\documentclass[aps, a4paper,superscriptaddress, nofootinbib, showpacs, 12pt]{revtex4}
\usepackage{amsmath,amssymb, bm, natbib}
\usepackage{dsfont}
\usepackage{epsfig}
\usepackage{graphicx}
\usepackage{slashed}
\usepackage{color}
\usepackage{ulem}
\usepackage{accents}
\usepackage{mathrsfs}
\usepackage{bigstrut}

\newcommand{\be}{\begin{equation}}
\newcommand{\ee}{\end{equation}}

\newcommand{\beq}{\begin{equation}}
\newcommand{\eeq}{\end{equation}}
\newcommand{\bea}{\begin{eqnarray}}
\newcommand{\eea}{\end{eqnarray}}
\newcommand{\ba}{\begin{align}}
\newcommand{\ea}{\end{align}}
\newcommand{\bfig}{\begin{figure}}
\newcommand{\efig}{\end{figure}}

\def\be{\begin{equation}}
\def\ee{\end{equation}}
\def\ba{\begin{eqnarray}}
\def\ea{\end{eqnarray}}
\def\br{\begin{array}}
\def\er{\end{array}}

\newcommand{\RaiseBrace}[1]{\raise1.5pt\hbox{$\displaystyle#1$}}

\begin{document}
\title{ High Scale Mixing Unification for Dirac Neutrinos}
\author{Gauhar Abbas}
\email{Gauhar.Abbas@ific.uv.es}
\affiliation{The Institute of Mathematical Sciences, Chennai 600 113, India.}
\author{Saurabh Gupta\footnote{Present address: Instituto de F\'\i sica, Universidade de S\~ao Paulo,
C. Postal 66318,  05314-970 S\~ao Paulo, SP, Brazil }}
\email{saurabh@if.usp.br}
\affiliation{The Institute of Mathematical Sciences, Chennai 600 113, India.}
\author{G. Rajasekaran}
\email{graj@imsc.res.in}
\affiliation{The Institute of Mathematical Sciences, Chennai 600 113, India.}
\affiliation{Chennai Mathematical Institute, Siruseri 603 103, India.} 
\author{Rahul Srivastava}
\email{rahuls@imsc.res.in}
\affiliation{The Institute of Mathematical Sciences, Chennai 600 113, India.}


\begin{abstract}

Starting with high scale mixing unification hypothesis, we investigate the renormalization group evolution of mixing parameters and masses for Dirac type neutrinos. Following this hypothesis, the PMNS mixing angles and phase are taken to be 
identical to the CKM ones at a unifying high scale. Then, they are evolved to a low scale using renormalization-group equations. The  notable feature of this hypothesis is that renormalization group evolution with quasi-degenerate mass pattern can explain largeness of leptonic mixing angles even for Dirac neutrinos. The renormalization group evolution ``naturally'' results in a non-zero and small value of leptonic mixing angle $\theta_{13}$. One of the important predictions of this work is that the mixing angle $\theta_{23}$ is non-maximal and lies only in the second octant. We also derive constraints on the allowed parameter range for the SUSY breaking and unification scales for which this hypothesis works.  The results are novel and can be tested by present and future experiments.  
 
\end{abstract}


\maketitle


One of the most important open questions in neutrino physics is whether neutrinos are Dirac or Majorana particles. There are dedicated ongoing experiments with the sole objective to determine the nature of neutrinos
\cite{Agostini:2013mzu,Auger:2012ar,Gando:2012zm,Alessandria:2011rc}. Answering this question is  essential in finding the underlying theory of neutrino masses and mixing. 
From theoretical perspective, Majorana neutrinos provide an elegant explanation for the observed smallness of neutrino masses through the celebrated sea-saw mechanism \cite{Minkowski,GellMann:1980vs,Yanagida:1979as,Glashow:1979nm,Mohapatra:1979ia}. 
However, even for the Dirac neutrinos, there exist a number of appealing models which can explain the smallness of neutrino masses. The smallness of the masses in these models is explained in various ways such as from gauged $B-L$ symmetry, by using extra heavy degrees of freedom, from K\"ahler potential of supergravity, from Grand Unified Theory (GUT) or compactification scales etc. \cite{Ma:2014qra, Ma:2015raa, Mohapatra:1986bd,ArkaniHamed:2000bq,Borzumati:2000mc,Kitano:2002px,Abel:2004tt}. For further details on alternatives to see-saw mechanism, see \cite{Mohapatra:2004vr,Smirnov:2004hs,Murayama:2004me}.

Furthermore, from the cosmological perspective there is no compelling reason to prefer Majorana neutrinos over Dirac neutrinos. For example, Dirac neutrinos can also provide a satisfactory explanation of the observed baryon asymmetry 
\cite{Dick:1999je,Murayama:2002je}. Therefore, Dirac neutrinos are as plausible as Majorana ones and only experiments can settle this issue. The various experiments \cite{Agostini:2013mzu,Auger:2012ar,Gando:2012zm,Alessandria:2011rc} 
looking for neutrinoless double beta decay have not seen any signal so far. Hence, in view of above considerations it is important to investigate various possible scenarios for Dirac neutrinos.

The neutrino oscillation parameters are the guiding light for model building. Present experimental scenario of neutrino physics is quite exciting due to recent measurement of the mixing angle  $\theta_{13}$, which is now established to be non-zero
\cite{Abe:2011sj,Adamson:2011qu,Abe:2012tg,Ahn:2012nd,An:2012eh}.   As a result of this rather precise measurement, many neutrino models are facing stringent constraints \cite{Antusch:2013ti}. Therefore, it is worthwhile to explore, in details, 
any scenario which ``naturally'' predicts non-zero value of $\theta_{13}$. One such promising scenario is the `High Scale Mixing Unification' (HSMU) hypothesis, which was proposed and studied in the context of Majorana 
neutrinos  \cite{Mohapatra:2003tw,Mohapatra:2005gs,Mohapatra:2005pw, Agarwalla:2006dj}. For a recent analysis see \cite{Abbas:2014ala, Srivastava:2015tza, Abbas_new}\footnote{ A similar approach can also be found in \cite{Haba:2012ar}.}. The central idea of this hypothesis is 
that the mixing parameters of the quark sector become identical to those of neutrino sector at some unification scale, which may be the GUT scale.

On theoretical side, in addition to the gauge coupling unification, GUT also unify quarks and leptons in a joint representation of the GUT symmetry group \cite{georgi,Fritzsch:1974nn}.  As a consequence, flavor structures of quark and 
lepton sectors are no longer disconnected.  This can lead to the relations between quark and lepton mixing angles \cite{Antusch:2011qg,Marzocca:2011dh}.  It is quite natural and appealing that HSMU may be a footprint of such an underlying GUT. 

In the earlier works on HSMU hypothesis, it has been shown that, if neutrinos are Majorana particles, have a normal hierarchy and quasi-degenerate mass spectrum, the HSMU hypothesis can explain the experimentally 
measured neutrino mixing parameters with minimal supersymmetric standard model (MSSM) as an extension of the standard model (SM) \cite{Mohapatra:2003tw,Mohapatra:2005gs,Mohapatra:2005pw, Agarwalla:2006dj}. At this point, we would like to 
point out that the inverted hierarchy of neutrino masses is not compatible with HSMU hypothesis, as it does not lead to the radiative magnification of the mixing angles \cite{Mohapatra:2003tw,Mohapatra:2005gs}. 

There are several reasons (both theoretical and experimental) which indicate that HSMU hypothesis is more natural for Dirac neutrinos than Majorana neutrinos. 
If neutrinos are Majorana particles, then the PMNS-matrix has 6 independent parameters; 3-mixing angles, 1-Dirac phase and 2-Majorana phases. On the other hand 
CKM-matrix has only 4 independent parameters; 3-mixing angles and 1-Dirac phase. Thus, there is a clear mismatch between number of parameters on two sides and hence a one-to-one correspondence is impossible.

In the previous works on HSMU for Majorana neutrinos (cf. \cite{Mohapatra:2003tw,Mohapatra:2005gs,Mohapatra:2005pw, Agarwalla:2006dj,Abbas:2014ala, Srivastava:2015tza, Abbas_new}), an extra ad hoc assumption for the initial value of Majorana phases had to be made. 
They were either fixed at zero or treated as free parameters. Since Majorana phases enter in the RG running of all other parameters \cite{Antusch:2003kp}, depending on their chosen values they can drastically change the RG running of other parameters. 
Hence, for Majorana neutrinos, the final results following from HSMU hypothesis depend strongly on the ad hoc choice of Majorana phases.  In contrast, the Dirac neutrino case is unambiguous as the CKM mixing parameters can be 
mapped in an exact one-to-one correspondence with the mixing angles and phase of the PMNS matrix at the unification scale.

Furthermore, the Majorana case faces additional difficulties in explaining the present neutrino oscillation data.  The RG running of $\Delta m^2_{sol}$ 
to low scales, leads to values larger than the present 3$\sigma$ range (cf. \cite{Mohapatra:2005gs,Abbas:2014ala, Srivastava:2015tza, Abbas_new} for details). Although, this mismatch can be explained by SUSY threshold corrections\footnote{If Majorana phases are assumed to satisfy 
a particular constraint equation, then there exist a parameter space where no threshold corrections are needed (see \cite{Agarwalla:2006dj, Abbas_new} for further details). } 
but such an explanation requires the ratio of sleptons masses to be in a particular range.
This is another ad hoc requirement that one has to impose for Majorana case in order to explain the present oscillation data. On other hand, in this letter, we show that the RG evolution in Dirac case satisfies 
the current neutrino oscillation data pretty well  without any threshold corrections.

In view of the above discussion, in this letter, we aim to investigate the HSMU hypothesis in a model independent setup, assuming that the nature of neutrinos is Dirac type and assay our predictions on the face of present experimental neutrino oscillation data.

The working of HSMU hypothesis requires MSSM as an extension of  SM and implemented in two steps.  In the first step, we follow bottom-up 
approach, where the CKM mixing angles ($\theta^q_{12}$, $\theta^q_{13}$, $\theta^q_{23}$) and the Dirac phase ($\delta^q_{CP}$) of quark sector \cite{Xing:2011aa} are evolved through SM renormalization-group (RG) equations \cite{Lindner:2005as} 
from a low scale ($M_Z$, mass of the $Z$ boson) to the supersymmetry (SUSY) breaking scale.  From the SUSY breaking scale to the unification scale, evolution is governed by MSSM RG equations \cite{Lindner:2005as}. After obtaining CKM mixing parameters 
at the unification scale, following HSMU hypothesis, we equate them to the PMNS mixing parameters at same scale, i.e. $\theta^{0,q}_{12}=\theta^0_{12}$, $\theta^{0,q}_{13} = \theta^0_{13}$, $\theta^{0,q}_{23} = \theta^0_{23}$ 
and $\delta^{0,q}_{CP} = \delta^0_{CP}$. The neutrino masses ($m^0_1$, $m^0_2$, $m^0_3$), at the unification scale, are taken as free parameters. Here, the superscript ``0'' denotes the corresponding values at the unification scale.

In the second step,  we follow top-down approach.  We evolve the mixing parameters and masses of neutrinos from the unification scale to the low scale. The running of mixing parameters and masses from the unification scale to SUSY 
breaking scale is governed by MSSM RG equations. From the SUSY breaking scale to low scale, the evolution
occurs through SM RG equations.  Keeping the present experimental status of SUSY searches  \cite{Craig:2013cxa} in mind, we have chosen the scale of SUSY breaking as $2$ TeV and the value of $\tan \beta$ to be $55$. 
Moreover, the unification scale is taken to be $2 \times 10^{16}$ GeV, the typical scale for GUTs. These values have been taken in most part of our work unless otherwise specified.

The RG equations for the running of mixing angles and masses are coupled partial differential equations \cite{Lindner:2005as} which cannot be solved analytically. However, the dominant contribution 
to the running of the mixing angles can be approximately given by 
\begin{eqnarray}
 \frac{d \theta_{12}}{dt}  & \propto & \frac{m^2}{\Delta m^2_{21}} \nonumber \\
 \frac{d \theta_{13}}{dt}, \, \frac{d \theta_{23}}{dt} & \propto & \frac{m^2}{\Delta m^2_{32}} 
 \label{ang}
\end{eqnarray}
where $t=\ln(\mu)$ and $\mu$ is the renormalization scale. Also, $m$ is the mean mass of neutrinos and $\Delta m^2_{ij} = m^2_i - m^2_j$ $(i,j = 1,2,3)$ represents the mass square differences. As is clear from (\ref{ang}), 
in order to achieve large angle magnification, the neutrino masses at unification scale should be chosen to be quasi-degenerate with normal hierarchy. As mentioned earlier, like Majorana case, in this case also, the inverted hierarchy for neutrino 
masses turns out to be incompatible with HSMU hypothesis. Furthermore, we choose the neutrino masses  such that all the oscillation parameters, at low scale, fall in 
the present experimental 3-$\sigma$ range obtained from the global analysis \cite{Capozzi:2013csa}. In our work, the RG evolution of masses and mixing parameters of quarks and neutrinos has been computed at two-loop level using a MATHEMATICA 
based package REAP  \cite{Antusch:2005gp}.  

It should be noted that there are only three input free parameters and we are fitting five experimentally known numbers as output parameters. Hence, it is an over-determined problem and there may be no solution at all.     
The fact that there exists a solution correctly fitting all the oscillation data is quite remarkable. Moreover, as we show later, the allowed HSMU scale is consistent with the GUT scale, despite not using the information about gauge coupling unification
at any place in our analysis.  This is quite significant and interesting in itself. This shows that perhaps there is some element of truth in the HSMU hypothesis.

\begin{figure}[h!tb]
 	\begin{center}\vspace{0.5cm}
 	 \includegraphics[width = 0.45 \textwidth]{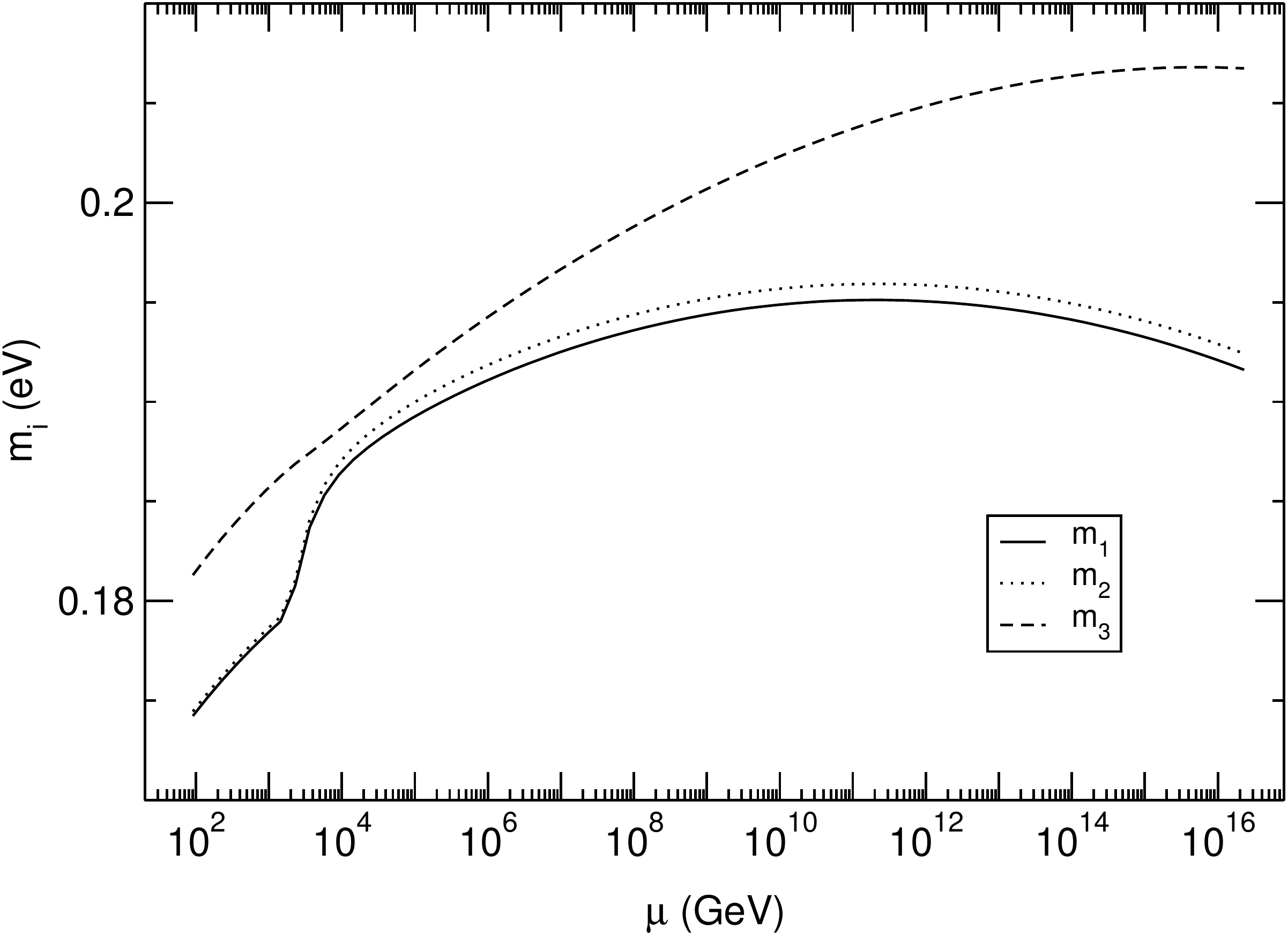} \hspace{0.2cm}
 	  \includegraphics[width = 0.45 \textwidth]{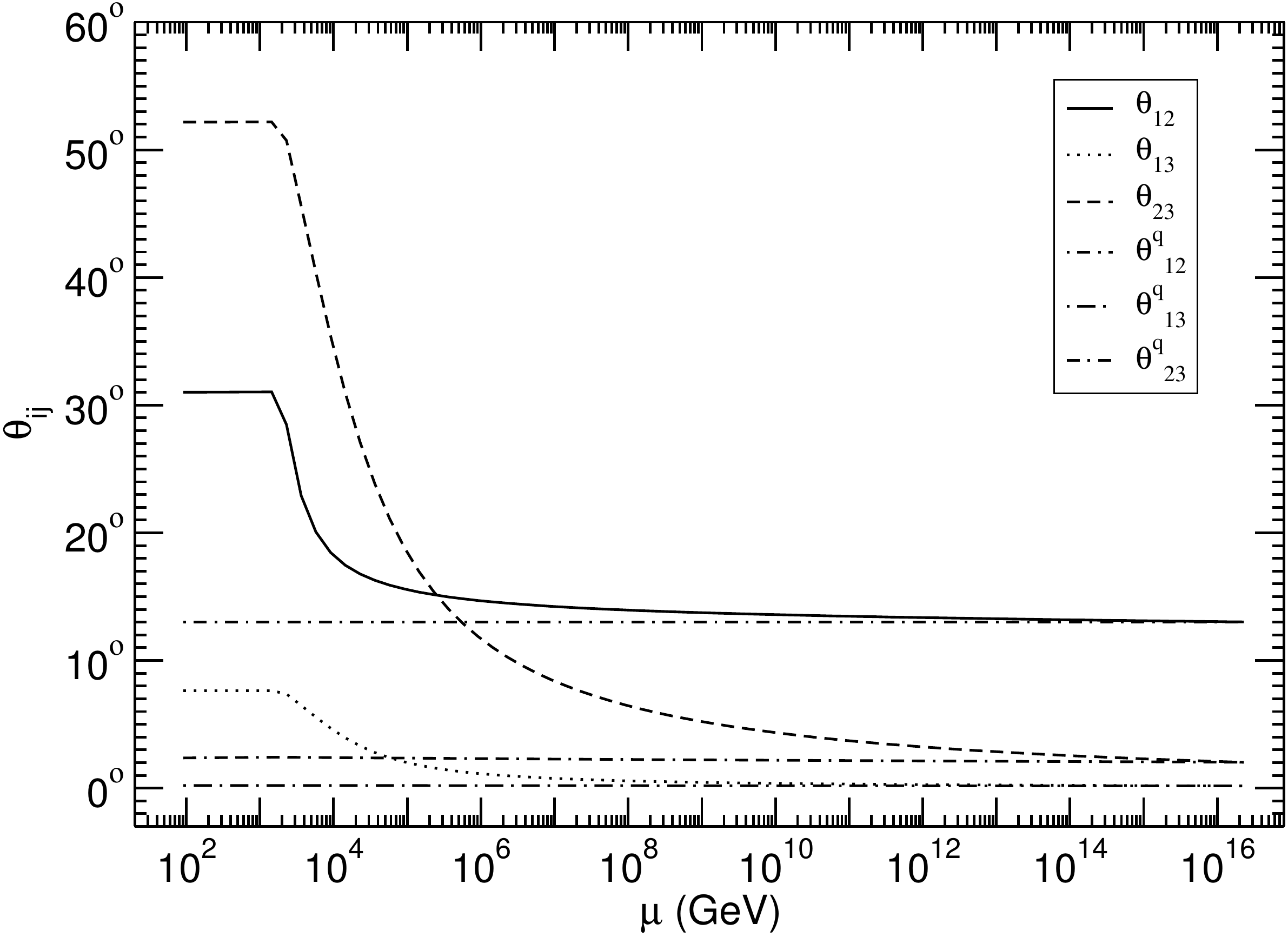} 
 	  \end{center}
 	    \begin{center}\vspace{0.5cm}
 	   \includegraphics[width = 0.45 \textwidth]{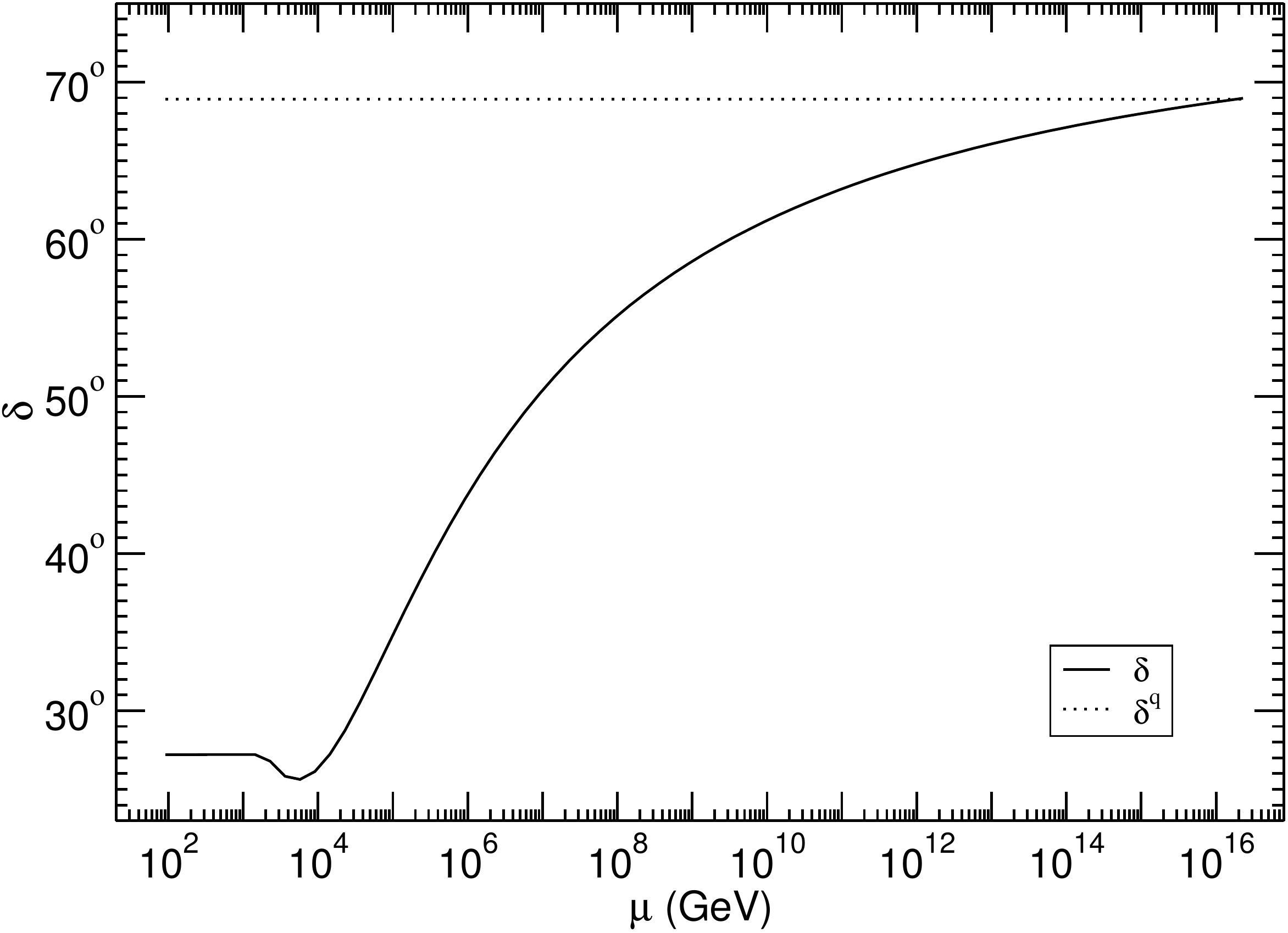} \hspace{0.2cm}
 	   \includegraphics[width = 0.45 \textwidth]{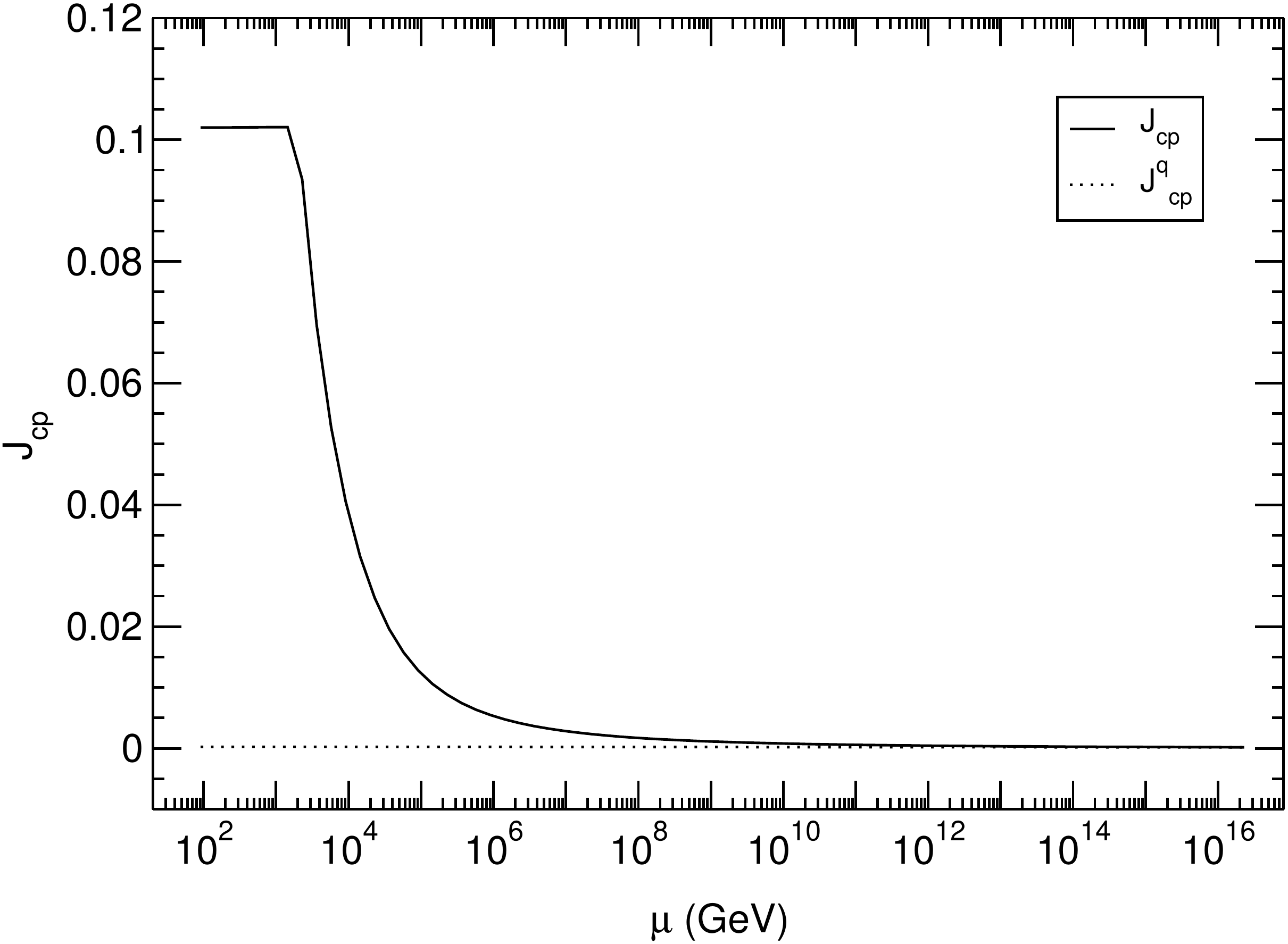}
	\caption{The RG evolution of the neutrino masses $m_i (i = 1,2,3)$, quark $\theta_{ij}^q$ $(i,j = 1,2,3)$ and leptonic mixing angles $\theta_{ij}$, neutrino Dirac phase $\delta$ and $J_{CP}$ with respect to the RG scale $\mu$, from the unification scale  ($2 \times 10^{16}$ GeV) to low scale ($M_Z$). The SUSY breaking scale and $\tan \beta$ are taken as $2$ TeV and 55, respectively. }
 	\label{fig1}
  	\end{center}\vspace{0.0cm}
 \end{figure}

The RG evolution of the neutrino masses, quark and neutrino mixing angles, neutrino Dirac phase and $J_{CP}$ from the unification scale ($2\times 10^{16}$ GeV) to low scale ($M_Z$), for certain typical values, is shown in Figure \ref{fig1}. 
In plotting Figure \ref{fig1}, the values of quark mixing parameters at unification scale, obtained from bottom-up running, are: $\theta_{12}^{0,q} = 13.02^\circ$, $\theta_{13}^{0,q} = 0.17^\circ$, $\theta_{23}^{0,q} = 2.03^\circ$ and $\delta_{CP}^{0,q} = 68.93^\circ$.
According to the HSMU hypothesis, the neutrino mixing parameters at unification scale are taken to be same as those of quark mixing parameters. We choose neutrino mass $m_2^0 = 0.1925$ $eV$ and mass square differences 
$\Delta m^2_{21}= 3.113 \times 10^{-4}$ $eV^2$, $\Delta m^2_{32} = 5.689 \times 10^{-3}$ $eV^2$ at the unification scale. Starting with these initial conditions, we obtain following values of oscillation parameters, after top-down running, at low scale: 
$\theta_{12} = 31.01^\circ$, $\theta_{13} = 7.63^\circ$, $\theta_{23} = 52.17^\circ$, $\delta_{CP} = 27.21^\circ$, $m_2 = 0.1745$ eV, $\Delta m^2_{sol} =  7.817 \times 10^{-5}$ $eV^2$ and $\Delta m^2_{atm} = 2.436 \times 10^{-3}$ $eV^2$.
It is clear from above data that all the low scale oscillation parameters are within their 3$\sigma$ range. 

The sum of neutrino masses at low scale, corresponding to the above mentioned values, turns out to be $\sum m_i = 0.530$ eV. The recent cosmological upper limit, from Planck collaboration, on the sum of neutrino masses is $\sum m_\nu < 0.72$ eV (Planck TT+lowP)
\cite{Planck:2015xua}. The sum of masses obtained from our analysis  satisfies this limit. Finally, the ``averaged electron neutrino mass'' obtained from our analysis is $\langle m_e \rangle = 0.174$ eV which is slightly below the present reach of KATRIN experiment \cite{Drexlin:2013lha}. However, it may be of interest to future experiments \cite{Ferri:2013qwa,Gatti:2012ii}. A more detailed analysis of the allowed ranges for various parameters  will be presented in a future work \cite{AGRS}.  

As shown in Figure \ref{fig1}, owing to the hierarchical nature of quark masses, the quark mixing angles change very little between the two scales. In the neutrino sector, the masses have normal hierarchy pattern  and 
because of our choice of quasi-degenerate masses, large angle magnification occurs.  The small and non-zero value of $\theta_{13}$, at low scale, can be attributed to the smallness of the quark mixing angle $\theta^{0,q}_{13}$ 
which is taken as the initial value for the neutrino mixing angle $\theta^0_{13}$.

\begin{figure}[thb]
  	\begin{center}\vspace{0.5cm}
  	 \includegraphics[width = 10.0cm]{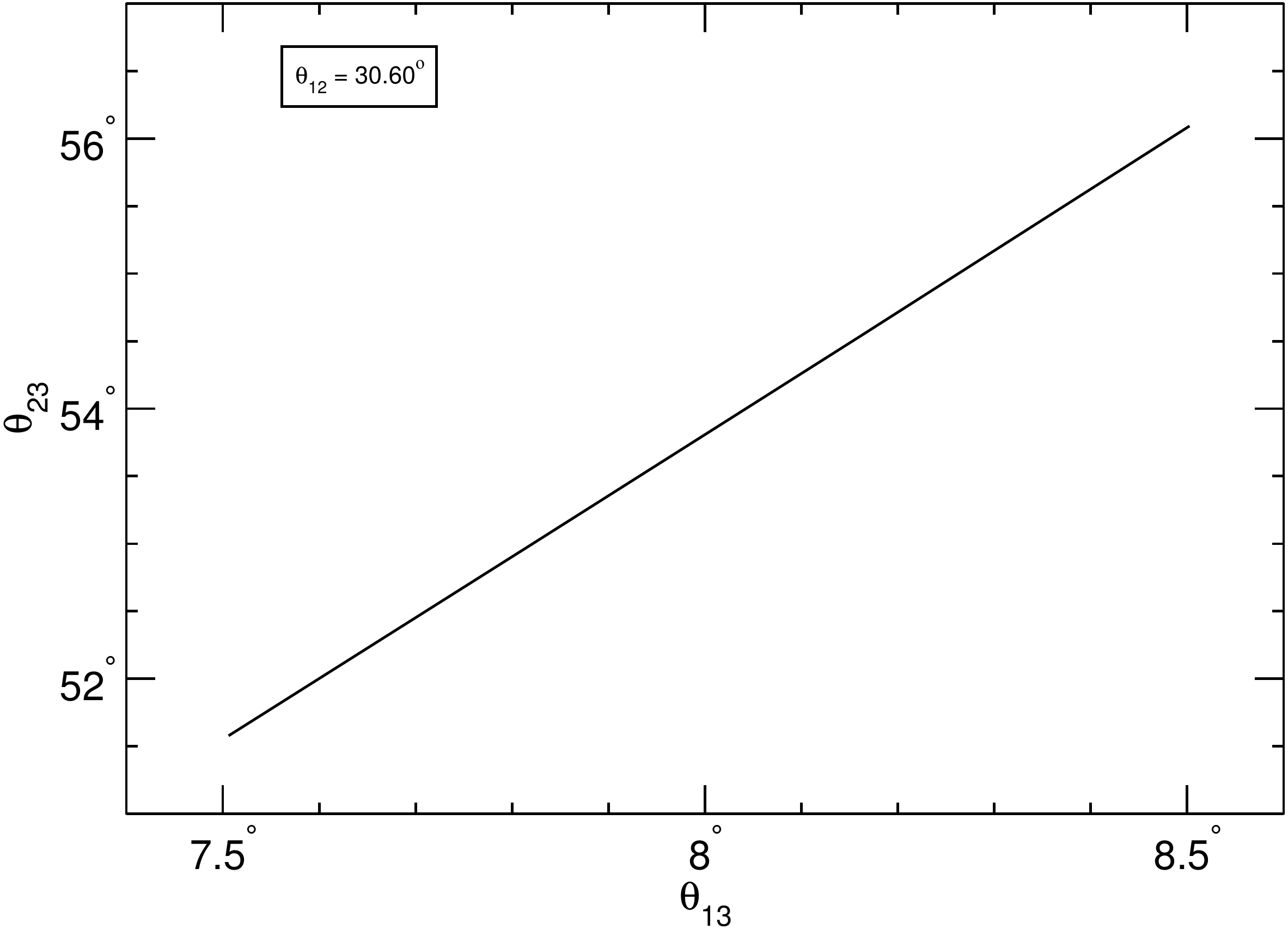}
 	\caption{The behaviour of leptonic mixing angle $\theta_{13}$ with $\theta_{23}$. In plotting this figure we have kept $\theta_{12}$ fixed at its lower end of the 3-$\sigma$ range and all other oscillation parameters within their 3-$\sigma$ range. }
	\label{fig2}
 	\end{center}\vspace{0.0cm}
\end{figure}

It is clear from (\ref{ang}) that the low scale values of mixing angles $\theta_{13}$ and $\theta_{23}$ are correlated as shown in Figure \ref{fig2}. In order to highlight this correlation, the value of $\theta_{12}$ at low scale is held fixed at $30.60^\circ$, which 
is near the lower end of allowed 3-$\sigma$ range. The purpose of this choice is only to illustrate our results. The effect of the variation of $\theta_{12}$ is small and will not change our conclusions. We observe that correlation between $\theta_{13}$ and $\theta_{23}$  evolves linearly.  For the present 3-$\sigma$ range of $\theta_{13}$ \cite{Capozzi:2013csa}, we find that $\theta_{23}$ lies only in the second octant  and towards the upper end of its 3-$\sigma$ range. 
This is a novel and robust result of this work. The effect of variation of all other parameters, including $\theta_{12}$,  do not change this prediction.   In addition to this, the non-zero value of $\theta_{13}$,
within its 3-$\sigma$ range, is a natural outcome of this analysis. The issue of full allowed parameter space for our predictions is the  subject of future study \cite{AGRS}.
Our predictions for $\theta_{13}$ and $\theta_{23}$ are unambiguous and can serve as an important test for the HSMU hypothesis. Various currently running and near future experiments such as INO, T2K, NO$\nu$A, LBNE, 
Hyper-K, PINGU \cite{Abe:2011ks,Patterson:2012zs,Adams:2013qkq,Ge:2013ffa,Kearns:2013lea,Athar:2006yb} have the potential to test our predictions.  

\begin{figure}[h!tb]
 	\begin{center}\vspace{0.5cm}
 	 \includegraphics[width = 10.0cm]{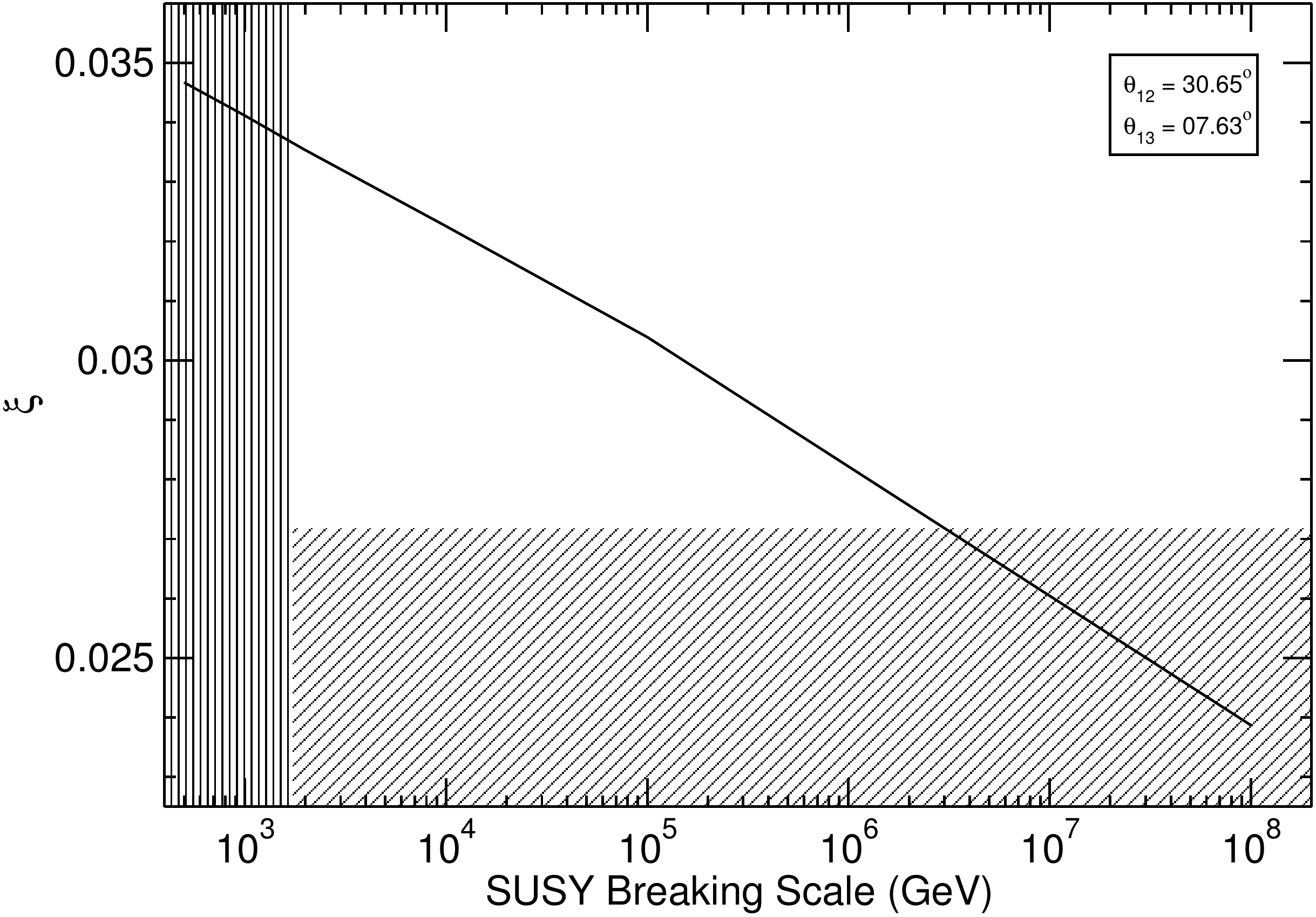}
	\caption{The variation of $\xi$ with SUSY breaking scale. The vertically shaded region is disallowed 
                    by LHC SUSY searches \cite{Craig:2013cxa} whereas the horizontal one lies outside the 3-$\sigma$ range of $\xi$ \cite{Capozzi:2013csa}. }
	\label{fig3}
 	\end{center}\vspace{0.0cm}
\end{figure}

So far in our analysis, we have kept the SUSY breaking and unification scales fixed at 2 TeV and $2 \times 10^{16}$ GeV, respectively. As mentioned before, there are only three input parameters which are used to determine five output parameters.
Therefore, it is not obvious that the HSMU hypothesis will work for any value of SUSY breaking and unification scales. Hence, it is important to address the issue of sensitivity of our 
analysis with respect to the choice of these parameters. Moreover, such an analysis also becomes important in the view of SUSY searches at LHC \cite{Craig:2013cxa}.

In Figure \ref{fig3}, we show the variation of SUSY breaking scale with respect to the parameter $\xi$, where $\xi = \Delta m_{sol}^2/\Delta m_{atm}^2$. The 3-$\sigma$ range of $\xi$ is $(2.72-3.72) \times 10^{-2}$ which is  
obtained from \cite{Capozzi:2013csa}. The parameter $\xi$ is taken because it provides the tightest constraints on the allowed parameter range for SUSY breaking scale. The values of $\xi$ lying in the unshaded region 
correspond to all the oscillation parameters at low scale being within their 3-$\sigma$ range. In plotting Figure \ref{fig3}, we have  kept $\theta_{12}$ and $\theta_{13}$ fixed,  at low scale, close to lower end of their 3-$\sigma$ range.  The lower bound on SUSY breaking scale comes from LHC searches which has excluded SUSY up to $1.7$ TeV in some scenarios \cite{Craig:2013cxa}.  The upper bound on SUSY breaking scale is obtained from the 3-$\sigma$ limit on $\xi$ which turns out to be around $4 \times 10^6$ GeV. Thus, keeping above constraints in mind, the SUSY breaking scale ranges from $1.7\times 10^3$ GeV to $4 \times 10^6$ GeV, within this framework. 
 
\begin{figure}[thb]
 	\begin{center}\vspace{1cm}
 	 \includegraphics[width = 10.0cm]{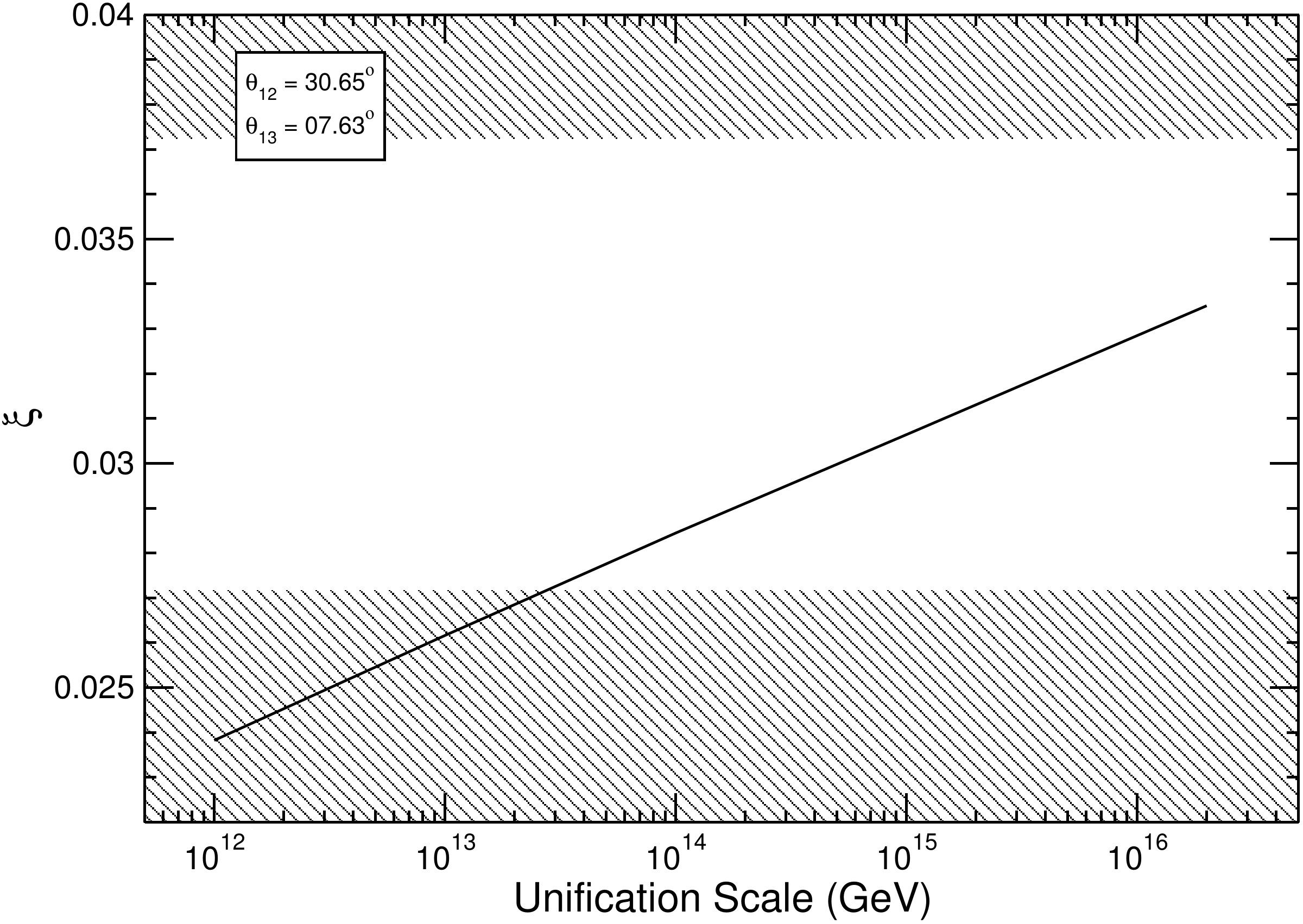}
	\caption{The variation of $\xi$ with unification scale. The shaded region lie outside the allowed parameter range of $\xi$ \cite{Capozzi:2013csa}.  }
	\label{fig4}
 	\end{center}\vspace{0.0cm}
\end{figure}

Similar analysis can also be performed for the variation of unification scale. In Figure \ref{fig4}, we show the variation of unification scale with respect to $\xi$.  We follow exactly same procedure as in Figure \ref{fig3}, except that, 
we have fixed SUSY breaking scale at $2$ TeV in this case.  The HSMU scale turns out to be consistent with the gauge coupling unification scale. In fact, as can be seen from Figure \ref{fig4}, the HSMU hypothesis is consistent with low scale neutrino oscillation data only if the HSMU scale is in and around the usual GUT scale.  We further observe that lower bound on  unification scale can be as low as $ 2 \times10^{13}$ GeV.

To conclude, at present, we do not know whether neutrinos are Dirac or Majorana particles. Only experiments can settle this long standing issue.  Keeping this spirit, in this work, we have investigated the HSMU hypothesis with Dirac type neutrinos. 
This hypothesis assumes that the mixing parameters of quark sector become identical to those of neutrino sector at some unification scale, which could be the GUT scale.  In addition to this, we also need MSSM as 
a natural extension of the SM.  After taking quark mixing parameters equal to the neutrino mixing parameters at unification scale, we run down the neutrino MSSM RG equations from unification scale to SUSY breaking scale.  From SUSY breaking 
scale to low scale $(M_Z)$, the running is governed by SM RG equations. We show that under HSMU hypothesis, RG evolution with quasi-degenerate masses can yield an explanation for largeness of leptonic mixing angles for Dirac neutrinos.  Although this fact is known for Majorana neutrinos \cite{Mohapatra:2003tw,Mohapatra:2005gs,Mohapatra:2005pw, Agarwalla:2006dj,Casas:2003kh}, it was not confirmed for Dirac neutrinos \cite{Lindner:2005as}. After RG evolution, we obtain all the neutrinos oscillation parameters, at low scale,  within their 3-$\sigma$ range \cite{Capozzi:2013csa}.

From this analysis, we have a clear and unambiguous prediction that $\theta_{23}$ is non-maximal and lies only in the second octant. This predictions is novel and robust. The precise determination of this angle is important to extract information 
about other oscillation parameters including $\delta_{CP}$. The range for $\theta_{13}$ is also tightly constrained. Moreover, the normal hierarchy and quasi-degeneracy of neutrino masses, essential to our analysis, can also 
provide an important test for this framework.  These predictions can be examined in present and the future experiments like INO, T2K, NO$\nu$A, LBNE, Hyper-K, PINGU \cite{Abe:2011ks,Patterson:2012zs,Adams:2013qkq,Ge:2013ffa,Kearns:2013lea,Athar:2006yb}.

We have also derived constraints on the allowed range of the SUSY breaking scale and unification scale.  The lower bound on SUSY breaking scale comes from the experimental SUSY searches \cite{Craig:2013cxa}. The upper bound on SUSY 
breaking scale, derived in this work, turns out to be around $ 4 \times 10^6$ GeV. Also, the unification scale for HSMU hypothesis can be taken anywhere from  the usual GUT scale down to about $ 2 \times10^{13}$ GeV.


\begin{acknowledgments}
RS would like to thank R.N. Mohapatra, E. Ma, J.W.F. Valle, R. Laha and A. Menon for their valuable comments and suggestions. 
The work of SG is supported in part by the CNPq, Brazil grant 151112/2014-2. 
\end{acknowledgments}


\end{document}